%% file: main.tex
\documentclass[runningheads]{llncs}
\usepackage{lscape}
\usepackage{algorithm}
\usepackage[noend]{algpseudocode}
\usepackage{amsmath}
\usepackage{graphicx}
\usepackage{todonotes}
\usepackage{caption}
\usepackage{multirow}
\usepackage{subcaption}

\usepackage{booktabs} % pretty tables
\usepackage{arydshln} % hdashline

\usepackage[utf8]{inputenc}        % req. for accent in Poincaré'
\usepackage{amsfonts}
\usepackage{commath} % Adds typesetting for differential equations and integral

\newcommand{\eg}{\emph{e.g., }}
\newcommand*\from{\colon}
\newcommand{\map}[3]{#1\from #2 \to #3}
\newcommand{\mapself}[2]{\map{#1}{#2}{#2}}
\newcommand{\bs}[1]{\boldsymbol{#1}}

\usepackage{xcolor}
\definecolor{webred}{rgb}{0.5,0,0}
\definecolor{webblue}{rgb}{0,0,0.8}
\usepackage[pdftex,colorlinks,citecolor=webblue,linkcolor=webred,]{hyperref}

\newtheorem{problemtheorem}{Problem}

\begin{document}
%
%\title{Verification of black-box dynamical systems via Koopman operator\thanks{Supported by organization x.}}
\title{Reachability of Black-Box Nonlinear Systems after Koopman Operator Linearization}
\titlerunning{Reachability of Black-Box Nonlinear Systems using Koopman Linearization}
\author{Stanley Bak\inst{1} \and
Sergiy Bogomolov\inst{2} \and
Parasara Sridhar Duggirala\inst{3}\and
Adam R. Gerlach\inst{4} \and
Kostiantyn Potomkin\inst{2}
}
\authorrunning{S. Bak et al.}
\institute{Department of Computer Science, Stony Brook University, NY, USA \and
School of Computing, Newcastle University, Newcastle Upon Tyne, UK
 \and
 Department of Computer Science, University of North Carolina at Chapel Hill, NC, USA
\and
Aerospace Systems Directorate, United State Air Force Research Laboratory, Wright–Patterson Air Force Base, OH, USA
}
\maketitle              % typeset the header of the contribution
\setcounter{footnote}{0}   % Required so, affiliations don't count for body footnotes

\input{sections/abstract}
\input{sections/intro}
\input{sections/preliminaries}

\input{sections/koopman}
\input{sections/algorithm}
\input{sections/eval}

\input{sections/related}
\input{sections/conclusions}
\input{sections/acks}
\bibliographystyle{splncs04}
\bibliography{bibliography,bak}
%
%
% \end{thebibliography}
\end{document}

%% file: sections/abstract.tex
\begin{abstract}

Reachability analysis of nonlinear dynamical systems is a
challenging and computationally expensive task.
Computing the reachable states for linear systems, in contrast,
can often be done efficiently in high dimensions.
In this paper, we explore verification methods that leverage
a connection between these two classes of systems based on
the concept of the Koopman operator.
The Koopman operator links the behaviors of a nonlinear system
to a linear system embedded in a higher dimensional space,
with an additional set of so-called observable variables.
Although, the new dynamical system has linear
differential equations, the set of initial states is defined with
nonlinear constraints.
For this reason, existing approaches for linear systems reachability cannot be used directly.
In this paper, we propose the first reachability algorithm that deals with this unexplored type of reachability problem.
Our evaluation examines several optimizations, and shows the proposed workflow is a promising avenue for verifying behaviors of nonlinear systems.

\keywords{Reachability Analysis  \and Nonlinear Systems \and Koopman Operator.}
\end{abstract}

%% file: sections/intro.tex
\section{Introduction}

Verification of cyber-physical systems requires algorithms that can reason over both software and the behavior of the physical world.
In particular, computing the post operation for a set of states is an important fundamental operation.
For software, the post operation computes how a set of variables is modified after executing a block of code.
For the physical world, the post operation determines how state variables can change after some amount of time has elapsed.
In hybrid systems terminology, the post operation is often called time-bounded reachability.
Given a set of initial states, some model of the physical system, and a time bound, the goal is to compute how the set of
states can change up to the time bound, possibly checking if some unsafe configuration is possible.

Reachability analysis for nonlinear systems is challenging, and is often the bottleneck for formal cyber-physical systems (CPS) analysis methods.
In this work we investigate nonlinear reachability approaches based on \emph{Koopman operator linearization}~\cite{koopman_book}.
Koopman operator linearization is a process where a nonlinear system can be approximated as a linear system with a large number of
so-called observable variables, each of which can be a nonlinear function of the original state variables.
This linear system can be computed either symbolically from differential equations or---importantly for black-box
systems---from data derived from real-world system executions or simulations~\cite{kutz_dynamic_2016}.
For reachability analysis, such a method is promising, as there exist highly-scalable methods to compute reachable sets for linear systems~\cite{girard2005hscc,leguernic09,bak2019hscc,bogomolov-et-al-2020:emsoft-2020}.
However, directly applying existing algorithms is not possible, as the observable variables are nonlinear
functions of the original state variables.
If the original initial set is described as a convex polytope, for example, the initial set of the Koopman linearized system may be
nonconvex, which current algorithms do not support.

To apply the Koopman method for reachability analysis, two issues must be overcome.
First, a Koopman linearized model of the nonlinear dynamics must be constructed whose behavior is a good approximation of the original system.
Second, linear reachability analysis methods must be modified to support nonlinear initial state sets.
The first problem is the focus of much current research on dynamical system analysis with the Koopman operator, and we do not focus on it here.
Instead, this paper's primary contribution is on the second problem, developing efficient algorithms for analysis for the class of systems
with linear dynamics and nonlinear initial sets.
We first show the problem can be solved using a nonlinear satisfiability module theory (SMT) solver to enforce the initial state constraints.
This Direct Encoding Algorithm is correct but may be slow in practice and even undecidable in theory.
We improve analysis efficiency through zonotope overapproximations of the nonlinear initial sets constructed using interval arithmetic,
as well as two abstraction-refinement techniques: (i) Hyperplane Backprogation and (ii) Zonotope Domain Splitting.

This paper is organized as follows.
First, we provide a brief review of Koopman operator linearization in Section~\ref{sec:koopman}.
Next, we describe our proposed verification algorithm and optimizations for Koopman linearized systems in Section~\ref{sec:algorithm}.
We then demonstrate the feasibility of the approach and evaluate each of the optimizations on a number of nonlinear systems in Section~\ref{sec:eval}.
We finish with related work and a conclusion.

%% file: sections/preliminaries.tex
\section{Preliminaries}
\label{sec:prelims}

Before detailing Koopman operator theory and our proposed reachability algorithm, we first introduce a few important mathematical preliminaries.

An interval $[a,b]$ (with $a \leq b$) denotes the set of numbers $x$ such that $a \leq x \leq b$. Arithmetic operations such as addition and multiplication over intervals $[a_1, b_1]$ and $[a_2, b_2]$ are defined using \emph{interval arithmetic}~\cite{interval_analysis_book} as follows.
\begin{align*}
  [a_1, b_1] + [a_2, b_2] &= [a_1+a_2, b_1+b_2]\\
  [a_1, b_1] \times [a_2, b_2] &= [\mathsf{min}\{a_1 a_2, a_1 b_2, b_1  a_2, b_1 b_2\}, \mathsf{max}\{a_1 a_2, a_1 b_2, b_1 a_2, b_1 b_2\}]
\end{align*}

The system under consideration evolves in the state space $\mathbb{R}^{n}$. Elements in the state space are denoted as ${\bs x}$, and scalars are denoted as $x$. Dynamical systems we consider in this paper evolve according to the differential equation:
\begin{equation}
  \label{eq:sys}
  \od{\bs x}{t} = \bs{f}({\bs x}).
\end{equation}

Often, $\bs{f}$ is a nonlinear function of the state ${\bs x}$.
We refer to such systems as non-linear dynamical systems. A trajectory of such nonlinear dynamical system described in Equation~\ref{eq:sys} is a function $\xi_{\bs{f}}({\bs x_0}, t)$, the solution to the differential equation $\bs{f}$, that takes as input an initial state ${\bs x_0}$ and time $t$ and returns the state of the system after time $t$.
We often drop the subscript $\bs{f}$ from the solution function $\xi_{\bs{f}}$ when it is clear from context.

When the function $\bs{f}({\bs x})$ is a linear function, i.e., $\bs{f}({\bs x}) = A {\bs x}$ where $A \in \mathbb{R}^{n \times n}$, the trajectory has a closed-form solution. More specifically, $\xi({\bs x_0},t) = e^{At}{\bs x_0}$ where $e^{At}$ is the matrix exponential that can be defined as:
$$
e^{At} = I + \frac{At}{1!} + \frac{(At)^2}{2!} + \frac{(At)^3}{3!} + \ldots
$$
% 
% Given the set of states represented by halfplane constraint $c^{T}x \leq d$. 
%
For the dynamical system $\dot{x} = Ax$, the set of states that reach $c^{T}x \leq d$ at time $t$ is given by the set of states $((e^{At})^{T}c)^Tx \leq d$.
That is, the backward propagation of a constraint $c^{T}x \leq d$ is obtained by right multiplying $c$ with $(e^{At})^T$.
We label this as \textbf{hyperplane backpropagation}.

An important part of Koopman linearization is the notion of observable functions, or observables, which are scalar functions from system's state space, $g: \mathbb{R}^{n} \rightarrow \mathbb{R}$. A vector of $k$ scalar-valued observables maps the state space to $\mathbb{R}^{k}$
%($k$ may be $\infty$)
and is denoted as $\bs{g}: \mathbb{R}^{n} \rightarrow \mathbb{R}^{k}$. Sometimes, we denote the observables as $\bs{y} = \bs{g}({\bs x})$.

In this paper we focus on safety verification at \emph{discrete} time instances with some step size $h > 0$, where
$h$ is assumed to evenly divide the analysis time bound $T$.
This both simplifies the problem and allows us to generate counterexamples demonstrating safety violations.
A trajectory on a nonlinear system can be safe in discrete time with respect to a given set of unsafe states.

\begin{definition}
  \label{def:safety}
Trajectory $\xi({\bs x_0}, t)$ is \textbf{discrete time safe} if and only if $\forall_{ 0 \leq i \leq T / h}$
$\xi({\bs x_0}, ih) \notin U$, where $U$ is an unsafe state set and $i \in \mathbb{Z}_{\geq 0}$ is the step number.
\end{definition}

The main problem we want to solve is a discrete-time bounded nonlinear verification problem.

\begin{problemtheorem}
  \label{prob:nonlinear}
  Given a set of initial states $I$, time step $h$, time bound $T$, and nonlinear dynamics $\bs{f}$,
  we want to prove the system is discrete-time safe, meaning that $\xi({\bs x_0}, t)$ is discrete-time safe for every initial state ${\bs x_0} \in I$.
\end{problemtheorem}

We assume $I$ and $U$ are given as as conjunctions of linear constraints, unless indicated otherwise.

%% file: sections/koopman.tex
\section{Koopman Operator Linearization}
\label{sec:koopman}
The modern study of dynamical systems is driven by Poincaré's \emph{state space} view of the underlying system. By considering the evolution of points in a state space, this view enables intuitive tools for analyzing, designing, controlling, and verifying dynamical systems. However, it can be ill-suited for certain classes of problems such as
uncertain systems~\cite{budisic_applied_2012,gerlach_koopman_2020} and systems without explicit equations describing their evolution (data-driven or black-box models)~\cite{jones_whither_2001}.

An alternative to this state space view is Koopman's \emph{observable space} view of dynamical systems. In contrast to the state space view, the observable space view considers the evolution of observables, or functions, of the given state space \cite{budisic_applied_2012} instead of the states themselves. This alternative view leads to the notion of the so-called Koopman operator \cite{koopman_hamiltonian_1931}. For dynamical system $\mapself{S}{\Omega}$ and observable $\map{g}{\Omega}{\mathbb{R}}$, the Koopman operator $\mathcal{K}$ is defined by
\begin{equation}
\mathcal{K}g = g\circ S, \qquad \forall g \in L^\infty \label{eq:koopman}
\end{equation}
As such, the Koopman operator is an infinite-dimensional linear operator on the space of scalar-valued functions of the state space \cite{koopman_hamiltonian_1931}. The spectral properties of this linear operator describe the evolutionary properties of the underlying dynamical system $S$, similar to finite-dimensional linear state space models (\eg eigen decomposition of a state matrix). However, unlike a finite-dimensional linear state matrix which describes the evolution of system states, the Koopman operator describes the evolution of scalar-valued functions as driven by the dynamics of the underlying system \cite{narasingam_koopman_2019}. 

As Eq.~\ref{eq:koopman} is equally valid for linear and nonlinear systems, the observable space view enables the linear treatment of full nonlinear dynamics via the Koopman operator. Thus, the Koopman operator has the potential to bridge nonlinear systems and existing linear tools for analysis, design, control, and verification \cite{brunton_koopman_2016,kutz_dynamic_2016} without sacrificing information, like with traditional linearization techniques \cite{budisic_applied_2012}.
Unfortunately, this theory comes with a practical cost because the Koopman operator is infinite-dimensional.
In theory, one can equivalently switch between representations of a dynamical system that are nonlinear finite-dimension (state space) or linear infinite-dimensional (observable space). In other words, one can \emph{lift} the nonlinear dynamics into a higher-, possibly infinite-, dimensional space where its dynamics are linear \cite{korda_linear_2018}.

For certain classes of systems, it is possible to obtain a finite-dimensional Koopman operator that describes the evolution of a system. In particular, these systems posses Koopman-invariant subspaces containing the system state \cite{brunton_koopman_2016}.

For example, consider the following continuous-time dynamical system:
\begin{equation}\label{eq:finite_koopman_nonlinear}
	\od{\bs x}{t}= \begin{bmatrix}
	\mu x_1 \\
	\lambda\del{x_2-x_1^4}
	\end{bmatrix}
\end{equation}
where $\bs x^\top = \sbr{x_1,x_2}$. If we consider the observables $g_1\del{\bs x}=x_1$, $g_2\del{\bs x}=x_2$, $g_3\del{\bs x}=x_1^4$, $g_4\del{\bs x}=x_1^3$, and $g_5\del{\bs x}=x_1^2$, the state can be lifted into a five-dimensional space where the system evolves linearly.  Using the relationships in Eq.~\ref{eq:finite_koopman_nonlinear}, the derivatives of the observables are linear, and can be written as
\begin{align}\label{eq:finite_koopman_linear}
	\od{}{t}\bs{g}\del{\bs x} &= \underbrace{\begin{bmatrix}
	\mu &0& 0& 0& 0\\
	0&\lambda&-\lambda&0&0\\
	0&0&0&4&0\\
	0&0&0&0&3\\
	2&0&0&0&0
	\end{bmatrix}}_{\tilde{\mathcal{K}}}\bs{g}\del{\bs x}
\end{align}
where $\bs{g}^\top=\sbr{g_1,g_2,g_3,g_4,g_5}$ and $\tilde{\mathcal{K}}$ is the infinitesimal Koopman operator \cite{lasota_chaos_2013}. The solution to this linear ordinary differential equation is then
\begin{align}\label{finite_koopman_solve}
   \bs{g}\del{\bs x_t}&=e^{\tilde{\mathcal{K}}t}\bs{g}\del{\bs x_0} \nonumber\\
                               &=\mathcal{K}_t\bs{g}\del{\bs x_0}
\end{align}
where $\mathcal{K}_t=e^{\tilde{K}t}$ is the Koopman operator as parameterized by $t$. 

As the system states are contained in the set of observables ($g_1$ and $g_2$) the states can be recovered from  Eq.~\ref{finite_koopman_solve} linearly as
\begin{equation}
	\bs x_t = \begin{bmatrix}
	1&0&0&0&0\\
	0&1&0&0&0
	\end{bmatrix}\mathcal{K}_t\bs{g}\del{\bs x}
\end{equation}

Unfortunately, identifying the observables that form a Koopman-invariant subspace of a system is a difficult problem and an active area of research.
However, simple algorithms exist for computing finite \emph{approximations} of the Koopman operator given time-series data.
These algorithms are primarily based on the Dynamic Mode Decomposition (DMD) algorithm \cite{schmid_dynamic_2010}, which performs regression of data to derive linear dynamics \cite{kutz_dynamic_2016}.

To fit a discrete-time linear system to data, the DMD algorithm starts by concatenating temporal snapshots of the system states $\bs x$ as columns of two data matrices, $X$ and $X'$. Given $m$ time instances of an $n$-dimensional system, these two $n\times m-1$ matrices are formed as
\begin{align}
	X&=\begin{bmatrix}\label{eq:DMD1}
	\bs x_1 & \bs x_2 & \ldots & \bs x_{m-1}
	\end{bmatrix} \\
	X'&=\begin{bmatrix}\label{eq:DMD2}
	\bs x_2 & \bs x_3 & \ldots & \bs x_{m}
	\end{bmatrix}
\end{align}
where $\bs x_k$ indicates the $k^\mathrm{th}$ temporal snapshot of $\bs x$. The best-fit least squares state transition matrix $A$ such that $X'\approx AX$ is then given by
\begin{equation}
	A=X'X^\dagger\label{eq:pinv}
\end{equation}
where $X^\dagger$ is the Moore-Penrose pseudoinverse of $X$. Because the DMD algorithm assumes linear dynamics, one can naturally extend this algorithm to approximate the linear Koopman operator from data by lifting the states in Eqs.~\ref{eq:DMD1} and \ref{eq:DMD2} using a finite set of observables called
a \emph{dictionary}. In other words, given some dictionary of $\ell$ pre-defined observables, where $\bs g\del{\bs x}=\sbr{g_1\del{\bs x},g_2\del{\bs x},\ldots,g_\ell\del{\bs x}}^\top$, lift Eqs.~\ref{eq:DMD1} and \ref{eq:DMD2} as 
\begin{align}
G&=\begin{bmatrix}\label{eq:obs_dmd1}
\bs g\del{\bs x_1} & \bs g\del{\bs x_2}  & \ldots & \bs g\del{\bs x_{m-1}} 
\end{bmatrix} \\
G'&=\begin{bmatrix}\label{eq:obs_dmd2}
\bs g\del{\bs x_2} & \bs g\del{\bs x_3}  & \ldots & \bs g\del{\bs x_{m}} 
\end{bmatrix}
\end{align}
and solve in the same manner as with DMD, leading to an approximation of the Koopman operator~\cite{williams_datadriven_2015}.

Similarly, one can approximate the infinitesimal Koopman operator by substituting data matrix $G'$ by  $\od{}{t}G$. Depending on the application at hand, these derivatives may be directly available via rate sensors or computable when synthetic data is leveraged. Alternatively, finite difference methods may be used. 

In practice it may be impractical to solve Eq.~\ref{eq:pinv} as presented, as the data matrices can be large and/or the data may be noisy.
In such cases, one can adjust the algorithm to use low-rank approximations of the data matrix $G$. This is typically achieved by truncating the non-dominate modes of the singular value decomposition (SVD) of $G$.
Koopman operator linearization for analysis and control are active areas of research in the applied mathematics community, with many further important details we could not review here for space reasons~\cite{mauroy_koopman_2020,kutz_dynamic_2016}.

For the purposes of this paper, we strive to use the Koopman observable space view to perform reachability analysis for nonlinear systems.
Note that the accuracy of this approach will depend heavily on the dictionary chosen during the DMD process, which is an ongoing research area.
As mentioned in the introduction, we do not focus on this problem here.
Instead we strive to develop reachability algorithms for the output of the Koopman linearization process.

%\textcolor{red}{Stan: Can we define problem 2 here similar to Problem 1 in in the preliminary section, that formally defines the problem the algorithm is solving (linear system with nonlinear observables)?}

%% file: sections/algorithm.tex
\section{Verifying Linear Systems with Nonlinear Observables}
\label{sec:algorithm}

Koopman operator linearization creates approximations to non-linear, possibly black-box systems.
The resulting systems have linear dynamics in the space of observables.
Reachability analysis of linear systems is a well-studied topic, and existing methods are efficient even with thousands or more state variables~\cite{bak2019hscc}.
However, the initial and unsafe constraints in the original problem are defined on the original system variables, not on the nonlinear observable variables.
This nonlinear relationship creates two additional tasks: (1) projection of initial set from state space variables to the space of observables, and (2) projection of reachable set in the space of observables back into the state space.
Problem (2) can be resolved by including the original state variables within the dictionary used during Koopman linearization.
This means that the projection from the space of observables to the original state variable can be written as a simple linear transformation, ${\bs x} = M \bs{g}({\bs x})$.
Problem (1), however, is more difficult to handle, and the main focus of the rest of this section.

\subsection{Direct Encoding of Nonlinear Constraints with SMT Solvers}
\label{sec:directdreal}
The first solution we propose, which we call the \emph{Direct Encoding Algorithm}, is to enforce the relationship between observables and state variables directly with nonlinear constraints.

Let the state space of the system to be $\bs{x}$ and the observables be $\bs{g}({\bs x})$.
Furthermore, the evolution of observables is determined by a linear differential equations, $\bs{g}({\bs x_t}) = \mathcal{K}\bs{g}({\bs x_0})$.
The state after time $t$, denoted as ${\bs x_t}$ is $M \bs{g}({\bs x_t})$.
Given an initial set $I$, and unsafe set $U$, the safety verification can be formulated as satisfiability of constraints in Equation~\ref{eq:koopmanSMT}.

\begin{align}
{\bs x_0} \in I, y = \bs{g}({\bs x_0}), \nonumber \\
y_t = \mathcal{K}_t y, \label{eq:koopmanSMT} \\
{\bs x_t} = M y_t, {\bs x_t} \in U. \nonumber
\end{align}

Given step size $h > 0$, for each time instant $t = i \times h$, an SMT solver can be invoked with the constraints in Equation~\ref{eq:koopmanSMT}.
Often, the initial set $I$ and unsafe set $U$ are specified as conjunctions of linear constraints.
For such cases, observe that the only nonlinear constraint in Equation~\ref{eq:koopmanSMT} is $y = \bs{g}({\bs x_0})$.
The complexity of finding an assignment of variables that satisfies these linear constaints dependd on the number and functions used for the observables.

For example, if the dictionary of observables are polynomials, SMT solvers can use algorithms like cylindrical algebraic decomposition~\cite{arnon1984cylindrical} that are theoretically guaranteed to terminate with a correct result.
In practice, this algorithm is doubly exponential in the number of variables, so a result may not be produced in a reasonable time.
If the dictionary includes transcendental functions like $\sin$ or $\cos$, the problem in general may not even be decidable.
In our implementation, we use the $\delta$-decidability SMT solver $\mathsf{dReal}$~\cite{dolzmann1997redlog,brown2003qepcad,gao2013dreal,de2008z3}, which theoretically always terminates but may produce an unknown result if the constraints are on the boundary of satisfiability (within a tolerance $\delta$), which we can then still flag as potentially unsafe.

Although the direct encoding works, for the reasons stated above it can be slow, so we next focus on optimizations and efficiency improvements.

\subsection{Overapproximating Nonlinear Constraints with Intervals}
\label{sec:interval}

Suppose that the initial set $I$ is given as hyperrectangles, as is often the case.
That is, $I=[x_{0l}^1; x_{0u}^1] \times \ldots \times [x_{0l}^n; x_{0u}^n]$.
Over such a domain, it is possible to compute conservative approximation of $y = \bs{g}({\bs x_0})$ where ${\bs x_0} \in I$ using interval arithmetic.
Alternatively, when $I$ is defined with linear constraints, we can compute upper and lower bounds on each of the variables using linear programming (LP) to construct box bounds, and then use those to construct the conservative approximation of $y = \bs{g}({\bs x_0})$.
Let the bounds we obtain on $y$ be such that $y \in [y_{l}^1, y_{u}^1] \times \ldots \times [y_{l}^k, y_{u}^k]$.
Substituting these in Equation~\ref{eq:koopmanSMT} results in the following constraints.

\begin{align}
y \in [y_{l}^1, y_{u}^1]\times \ldots \times [y_{l}^k, y_{u}^k] , \nonumber \\
y_t = \mathcal{K}_t y, \label{eq:koopmanInterval} \\
{\bs x_t} = M y_t, {\bs x_t} \in U. \nonumber
\end{align}

This set of constraints can be solved efficiently using LP, as is done in linear systems reachability analysis using zonotopes~\cite{girard2005hscc} or linear star sets~\cite{bak17hscc}.

While this method can be very efficient, the overapproximation of $\bs{g}({\bs x})$ using interval arithmetic might yield a very coarse overapproximation.
This would mean that spurious unsafe executions of the system may be found, due to the overapproximation.
To overcome this the \emph{Interval Encoding Algorithm} uses a hybrid approach, where an LP is first solved at each step according to Equation~\ref{eq:koopmanInterval}, and only if the LP is feasible will we then call the $\mathsf{dReal}$ SMT solver with the nonlinear constraints from Equation~\ref{eq:koopmanSMT}.

\subsection{Hyperplane Backpropagation}
\label{sec:hyback}

One of the building blocks that helps us improve efficiency
is the notion of \textbf{hyperplane backpropagation}.
Consider the unsafe set given as $U = [U^1_l; U^1_u] \times \ldots \times [U^n_l; U^n_u]$, observables $\bs{g}({\bs x})$, and ${\bs x} = M \bs{g}({\bs x})$.
Since we assumed the Koopman dictionary contained the original state variables, we can directly encode $U$ as constraints in the observable observable space, which we write as $\bs{g}(U)$.
If $q^{T}y \leq r$ is a halfplane constraint in $\bs{g}(U)$, then the corresponding constraint for propagating this constraint back by time $t$ is obtained by $(\mathcal{K}^{T}q)^T y \leq r$.
%
% If $y' \in \bs{g}(U)$, then the corresponding initial valuation of observables $y$ should be $\mathcal{K}^{-1}\bs{g}(U)$.
%
%\textcolor{red}{Stan: were we really doing an inverse? I thought we were doing something with transpose of the matrix exponential similar to reachability with support functions?}
%
We perform hyperplane backpropagation for all the constraints in $\bs{g}(U)$.
In the Koopman linerization literature, this operation is called \emph{pull-back} operation~\cite{meyers2019koopman}.
Any state that satisfies all the constraints obtained by performing the hyperplane backpropagation will end up in $\bs{g}(U)$ after time $t$.
%
% By performing the inverse transformation, we \emph{propagate} the \emph{constraints} specified as a part of unsafe set, into constraints on the initial set.
%
We label these constraints as $\mathsf{PropCons}(U)$.
While we have explicitly specified this only for unsafe sets that are intervals, this can be easily extended to unsafe sets specified as conjunction of half-spaces.
%
%In the remainder of this section, we present two main enhancements, namely, \emph{domain contraction}, and \emph{interval refinement} that use constraint propagation.

%
%In this subsection, we present an efficient approach to verify systems constructed via Koopman operator. Later, in the evaluation section, we demonstrate the efficiency of the presented approaches and discuss their performance.
%
%\begin{algorithm}
%\caption{Reachability algorithm}\label{alg:comb}
%\begin{algorithmic}[H]
%
%\Function{CheckDomain}{$A, I, U, t$}
%\State $back_{U} = U*e^{(A) t}$
%\State $inter_t = I \cap back_{U}$
%\If{$\!sempty(inter_t)$}
%\State $return \ \texttt{safe}$
%\EndIf
%\State $return \ dreal(A, I, U, t)$
%\EndFunction
%\end{algorithmic}
%\end{algorithm}
%
%\subsubsection{Combine techniques} First, we combine the two methods presented above (see Algorithm~\ref{alg:comb}).
%
%In particular, we first overapproximate an initial set with interval arithmetics as in Section~\ref{sec:interval}. Then, for each time moment, we check whether the intersection with an unsafe region is empty or not. Finally, if the intersection is not empty, we call dReal to look for counterexamples.
%% particular time moments with non-linear initial set

%\subsubsection{Domain Contraction}

We use the above process to backpropagate each of the unsafe set constraints into the initial set, i.e., $\mathsf{PropCons}(U)$.
We then compute the overlap between the projection of initial set into the observables ($\bs{g}(I)$) and compute its overlap with propagated constraints, i.e., $\bs{g}(I) \cap \mathsf{PropCons}(U)$, using interval arithmetic.
We then project this set back into the initial set as $M (\bs{g}(I) \cap \mathsf{PropCons}(U))$.
If this projected set $I'$ is a strict subset of initial set $I$, then, one can specify the possible violation of safety with higher precision.
We then repeat the process with the new set $I'$.
If, during this process $I'$ is empty, then, none of the trajectories go into the unsafe set; we can declare the system to be safe.
If $I'$ is same as the set $I$, then this approach does not further improve precision.
In this case, the \emph{Hyperplane Backpropagation Algorithm} would invoke $\mathsf{dReal}$ and instantiate the constraints given in Equation~\ref{eq:koopmanSMT} with $I'$ instead of $I$.
The pseudocode of the core step of the algorithm is provided in Algorithm~\ref{alg:domainContract}.

%We extend our previous algorithm with the co-called back refinement (see Algorithm~\ref{alg:backprop}). If the intersection between the unsafe region and the reachable set is not empty, we compute the intersection and then recalculate boundaries for the observables. In other words, once the intersection is computed, we take \textbf{only} the boundaries of the original system variables and utilize again interval arithmetics for the observables. We repeat this step until the new set is equal to the set before refinement.
% We add to the algorithm is to refine an initial set after checking whether intersection with unsafe region is empty. In particular, if intersection is not empty, we

%\begin{algorithm}
%\caption{Different refinement options}\label{alg:backprop}
%\begin{algorithmic}[H]
%\Function{RefineDomain}{$A, I, U, t, origin\_vars, observables\_funcs$}
%\State $cpa\_origins = projectOntoOrigins(I, origin\_vars)$
%
%\State $return \ getBoundaries(observables\_funcs, \ cpa\_origins)$
%\EndFunction
%
%\Function{CheckDomain}{$A, I, U, t, origin\_vars, observables\_funcs$}
%\State $back_{U} = U*e^{(A) t}$
%\State $inter_t = I \cap back_{U}$
%\If{$\!sempty(inter_t)$}
%\State $return \ \texttt{safe}$
%\EndIf
%\State $temp\_inter_t = EmptySet$
%\While{$temp\_inter_t != inter_t$}
%\State $temp\_inter_t = inter_t$
%\State $inter_t = RefineDomain(A, inter_t, U, t, origin\_vars, observables\_funcs) \cap back_{U}$
%\If{$\!sempty(inter_t)$}
%\State $return \ \texttt{safe}$
%\EndIf
%\EndWhile
%\State
%\State $return \ dreal(A, inter_t, U, t)$
%\EndFunction
%\end{algorithmic}
%\end{algorithm}

\begin{algorithm}
\caption{Hyperplane Backpropagation Algorithm}\label{alg:domainContract}
\begin{algorithmic}[H]
\Function{Backpropagate}{$I, U, t, {\bs x}, \bs{g}({\bs x}), \mathcal{K}$}
\State \textbf{Output:} A smaller initial set $I'$ after backpropagation, or $\emptyset$ if safe.

%\State $cpa\_origins = projectOntoOrigins(I, origin\_vars)$
%\State $return \ getBoundaries(observables\_funcs, \ cpa\_origins)$
%\EndFunction
%\Function{CheckDomain}{$A, I, U, t, origin\_vars, observables\_funcs$}

\State $U_{o} = \bs{g}(U)$
\State $\mathsf{PropCons} = \mathcal{K}^{-1} U_{o}$
\State $I_{o} = \bs{g}(I)$
\State $I' = M(I_{o} \cap \mathsf{PropCons})$
\While{$I' \subset I$}
\If{$I' = \emptyset$}
\State \textbf{return} \texttt{safe}.
\EndIf
\State $I_{o} = \bs{g}(I')$
\State $I' = M(I_{o} \cap \mathsf{PropCons})$
\EndWhile
\State \textbf{return} $I'$
\EndFunction
\end{algorithmic}
\end{algorithm}

\subsection{Zonotope Domain Splitting}

The reachable states at each time step encoded by Equation~\ref{eq:koopmanInterval}
can be represented using a zonotope~\cite{girard05}, where the box domain of the zonotope is the initial set $[y_{l}^1, y_{u}^1]\times \ldots \times [y_{l}^k, y_{u}^k]$.
The accuracy of this zonotope overapproximation can be improved by splitting each dimension's interval domain into several smaller subintervals and computing the new interval overapproximation in each of the subintervals.

For the proposed \emph{Zonotope Domain Splitting Algorithm}, we use a
heuristic that always splits the variable with the maximum range.
After splitting, we then perform hyperplane backpropagation on each of the smaller intervals.
This process can be repeated until either the overapproximation is safe for the current step, or some predetermined $\mathsf{upperLimit}$ is reached on the number of splits.
Upon reaching the limit, we then invoke $\mathsf{dReal}$ using the nonlinear constraints from Equation~\ref{eq:koopmanSMT}.
As a result, $\mathsf{dReal}$ is invoked with smaller initial sets, thus helping the numerical procedure to terminate faster.
However, as we show later in the evaluation, performing refinement too often can harm the performance of the verification procedure, as splitting can increase the number of queries needed.
The pseudocode for the core of the zonotope domain splitting algorithm is given in Algorithm~\ref{alg:binRefine}

%We also introduce a binary refinement (see Algorithm~\ref{alg:bin}) to reduce the size of the sets passed into non-linear solvers and verify smaller `boxes' with reduced approximation error. For this purpose, at each level of binary refinement we pick the `widest' interval out of our original system variables and split it half into two equal hyperrectangulars. We repeat binary refinement until we reach some $max\_level$ number of binary refinements. Later, in the experimental section, we showcase that large number of binary refinements can harm the performance of the overall algorithm.

%\begin{algorithm}
%\caption{Different refinement options}\label{alg:bin}
%\begin{algorithmic}[H]
%
%\Function{RefineAndVerify}{$I, U,  {\bs x}, \bs{g}({\bs x}), \mathcal{K}, t, level$}
%\If{$level \leq max\_level$}
%\State $temp\_bin_1, temp\_bin_1 = split(I)$
%\State $level += 1$
%\State $temp\_bin_1 = RefineDomain(A, temp\_bin_1, U, t, origin\_vars, observables\_funcs) \cap back_{U}$
%\State $temp\_bin_2 = RefineDomain(A, temp\_bin_2, U, t, origin\_vars, observables\_funcs) \cap back_{U}$
%\State return CheckDomain($A, temp\_bin_1, U, t, origin\_vars, observables\_funcs, level$) \&\& CheckDomain($A, temp\_bin_2, U, t, origin\_vars, observables\_funcs, level$)
%\EndIf
%
%\State $return \ dreal(A, inter_t, U, t)$
%\EndFunction
%\end{algorithmic}
%\end{algorithm}

\begin{algorithm}
\caption{Zonotope Domain Splitting Algorithm}\label{alg:binRefine}
\begin{algorithmic}[H]

\Function{ZonoSplitting}{$I, U,  {\bs x}, \bs{g}({\bs x}), \mathcal{K}_t, level$}
\If{$level \leq max\_level$}
\State $I_1, I_2 = split(I)$
\State $I_1' = Backpropagate(I_1, U, {\bs x}, \bs{g}({\bs x}), \mathcal{K}_t)$
\State $I_2' = Backpropagate(I_2, U, {\bs x}, \bs{g}({\bs x}), \mathcal{K}_t)$
\State \texttt{safe1} = $(I_1' = \emptyset)$ \textbf{or} ZonoSplitting($I_1', U,  {\bs x}, \bs{g}({\bs x}),
\mathcal{K}_t, level + 1$)
\State \texttt{safe2} = $(I_2' = \emptyset)$ \textbf{or} ZonoSplitting($I_2', U,  {\bs x}, \bs{g}({\bs x}),
\mathcal{K}_t, level + 1$)
\If{\texttt{safe1} \textbf{and} \texttt{safe2}}
\State \textbf{return} \texttt{UNSAT}
\Else
\State \textbf{return} \texttt{SAT}
\EndIf
\Else
\State \textbf{return} $\mathsf{dReal}(I, U, {\bs x}, \bs{g}({\bs x}), \mathcal{K}_t)$
\EndIf
\EndFunction
\end{algorithmic}
\end{algorithm}

Notice that the order of applying hyperplane backpropagation and zonotope domain contraction can alter the performance of the verification procedure.
While we prefer performing hyperplane backpropagation at each iteration, delaying the process until the interval under consideration becomes smaller could be a useful heuristic for some examples.
In our evaluation, we consider various possible combinations of these two methods.
In our experience, invoking \textsf{dReal} with the smaller initial sets succeeds to either prove safety or generate counterexample quicker than larger initial sets.

%\subsubsection{Whole algorithm} Finally, we combine all the ideas presented above into one full algorithm. First, we apply back refinement. Then, once we cannot  further refine we call binary refinement up to defined $max\_level$. If even after binary refinement we cannot verify whether the system is safe, we call the non-linear SMT solver (dReal in our case).

%% file: sections/eval.tex
\section{Evaluation}
\label{sec:eval}

In this section, we present experimental results of using the Koopman operator in reachability analysis.
We implemented our algorithms in Julia, using the LazySets package of JuliaReach~\cite{bogomolov2019juliareach}, the PyCall.jl\footnote{\url{https://github.com/JuliaPy/PyCall.jl}}
%\todo{ this doesn't have a proper citation. What is standard for this conf.? To cite software URL or to include as footnote? DataDrivenDiffEq need updated to match}
package to call $\mathsf{dReal}$~\cite{gao2013dreal}, the DataDrivenDiffEq.jl\footnote{\url{https://datadriven.sciml.ai/}}
package for Koopman operator linearization via Extended DMD and the DifferentialEquations.jl~\cite{rackauckas_differentialequationsjl_2017} package to generate numerical simulations. A Sobol sequence in the set of initial conditions is used to determine the initial conditions for the simulations. The resulting data matrices for each simulation are then combined via column-wise concatenation per \cite{tu_dynamic_2014}.

For each system, the dictionary of observables for Koopman linearization included multivariate polynomial basis functions up to a fixed order for
the original state variables, $\sin t$ and $\cos t$, and combinations of these (\eg $\bs x \sin^a t \cos^b t$).
Lastly, SVD truncation was performed as described in Section~\ref{sec:koopman} to remove any non-dominant modes.

Note that although we know the nonlinear differential equations for each system, we only used simulation data to perform Koopman linearization, treating each system as a black box.
Since few approaches can work directly with black-box systems, this allows us to compare against other reachability methods that require knowing the system's dynamics.

%Dominate modes are determined visually by looking for outliers in plots of the singular values.

%\centering\begin{tabular}{|r|c|c|c|}
%	\hline
%	& State Space Poly.~Basis & $\sin$/$\cos$ Poly.~Basis \\
%	\hline
%	Roessler Model&  &  \\
%	\hline
%	Steam Model&  &   \\
%	\hline
%	Coupled Van der Pol Oscillator&  &   \\
%	\hline
%	Biological Model&  &   \\
%	\hline
%\end{tabular}

\input{sections/systems}

\input{sections/results}

%% file: sections/systems.tex
\subsection{System Definitions}
%\textcolor{red}{Note: If we need space, much of this part can be moved to an online appendix.}

We evaluate our algorithms on four benchmark nonlinear systems.
As our algorithms are sensitive to the distance between the reachable set and the unsafe region, we consider parameterized unsafe regions, where a parameter $i$ controls the distance of the reachable set to the unsafe region. The nonlinear differential equations for each system are available on HyPro~\cite{schupp2017h} benchmark website\footnote{\url{https://ths.rwth-aachen.de/research/projects/hypro/benchmarks-of-continuous-and-hybrid-systems/}}.

\subsubsection{Roessler model}

% The Roessler attractor is modeled by the following ODE:

% \begin{align*}
%     \dot{x} &= -y -z\\
%     \dot{y}
%     &=x+0.2\cdot y\\
%     \dot{z}&=0.2+z\cdot(x-5.7)
% \end{align*}
We run the Roessler attractor\footnote{\url{https://ths.rwth-aachen.de/research/projects/hypro/roessler-attractor/}} with the following initial set: $x(0) \in [-0.05,0.05], y(0) \in [-8.45,-8.35], z(0) \in [-0.05,0.05]$.
 In our experiments we use the following parameterized unsafe region: $y\geq 6.375 - 0.025 \cdot i$, where $i \in [0, 20]$.
 The Koopman linearized system contains 70 observables.

\subsubsection{Steam model}

\begin{figure}[t]
% \begin{subfigure}{.5\textwidth}
%     \centering
%     \includegraphics[width=0.7\columnwidth]{figs/sims/roesslerNew.png}
%     \caption{Roessler model.}
%     \label{fig:roesslerSim}
% \end{subfigure}%
\begin{subfigure}{.5\textwidth}
    \centering
    \includegraphics[width=1.\columnwidth]{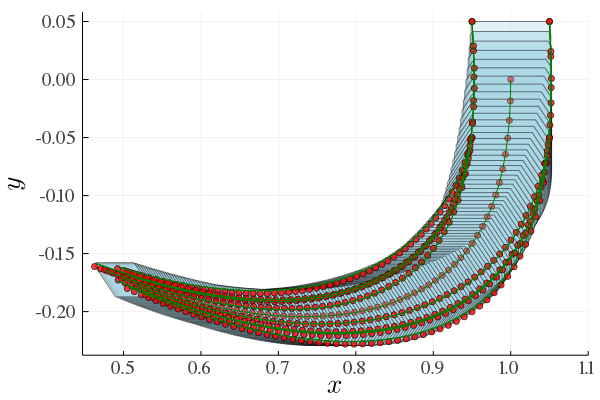}
    \caption{Steam model.}
    \label{fig:steamSim}
\end{subfigure}
% \begin{subfigure}{.5\textwidth}
%     \centering
%     \includegraphics[width=0.7\columnwidth]{figs/sims/coupleVP.png}
%     \caption{Coupled Van der Pol oscillator.}
%     \label{fig:coupleVPsim}
% \end{subfigure}%
\begin{subfigure}{.5\textwidth}
    \centering
    \includegraphics[width=1.\columnwidth]{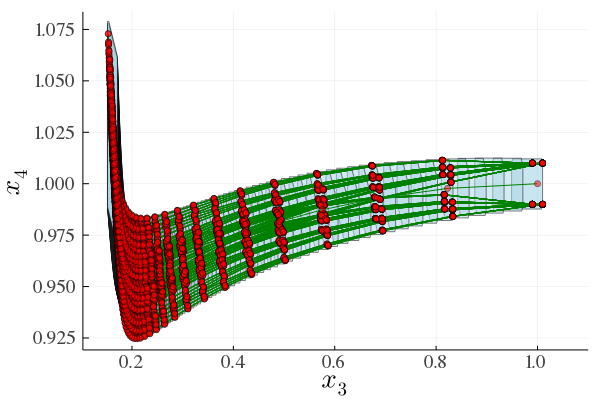}
    \caption{Biological model.}
    \label{fig:bioModelSim}
\end{subfigure}
\caption{Simulations of the original non-linear dynamical systems (green), their approximations obtained via Koopman operator (red) and Taylor Model approximations via Flow* (blue), show that the Koopman linearized model visually produces good approximations to the behaviors of the nonlinear system.}
\label{fig:sims}
\end{figure}

% The steam governor system~\cite{sotomayor2007bifurcation} is modeled by the following ODE:
% \begin{align*}
%     \dot{x} &= y\\
%     \dot{y}
%     &=z^2\cdot \sin(x)\cdot \cos(x)-3 \cdot y\\
%     \dot{z}&=\cos(x) - 1
% \end{align*}
The steam governor~\cite{sotomayor2007bifurcation} system \footnote{\url{https://ths.rwth-aachen.de/research/projects/hypro/steam-governor/}} is modeled with the following initial set: $x(0) \in [0.95,1.05], y(0) \in [-0.05,0.05], z(0) \in [0.95,1.05]$. In our experiments we use the unsafe region $y\leq -0.25 + 0.01 \cdot i$, where $i \in [0, 10]$. The Koopman linearized system contains 71 observables. Figure~\ref{fig:steamSim} compares simulations and reachable sets of the original nonlinear system with the result of Koopman linearization.

% \begin{figure}[h!]
%     \centering
%     \includegraphics[width=0.7\columnwidth]{figs/evals/steam.png}
%     \caption{Steam data}
%     \label{fig:steamSim}
% \end{figure}
%
% \begin{figure}[h!]
%     \centering
%     \includegraphics[width=0.7\columnwidth]{figs/evals/steamD.png}
%     \caption{Steam error}
%     \label{fig:steamSim}
% \end{figure}

\subsubsection{Coupled Van der Pol oscillator}

% The composed model of two coupled Van der Pol oscillators~\cite{rand1980bifurcation} system  is modeled by the following ODE:
% \begin{align*}
%     \dot{x_1} &= y_1\\
%     \dot{y_1}
%     &=(1-x^2_1)\cdot y_1 - x_1 + (x_2-x_1)\\
%     \dot{x_2}&=y_2\\
%     \dot{y_2}&=(1-x^2_2)\cdot y_2-x_2+(x_1-x_2)
% \end{align*}
 The composed model of two coupled Van der Pol oscillators~\cite{rand1980bifurcation} system \footnote{\url{https://ths.rwth-aachen.de/research/projects/hypro/coupled-van-der-pol-oscillator/}} is modeled with the following initial set: $x_1(0) \in [-0.025,0.025], y_1(0) \in [4.975,5.025], x_2(0) \in [-0.025,0.025], y_2(0) \in [4.975,5.025]$. In our experiments we use the following unsafe region: $x\geq 1.25 - 0.05 \cdot i$, where $i \in [1,16]$ . The linearized system contains 131 observables.

%
% \begin{figure}[h!]
%     \centering
%     \includegraphics[width=0.7\columnwidth]{figs/evals/coupleVP.png}
%     \caption{Couple vp data}
%     \label{fig:coupleVPsim}
% \end{figure}
%
% \begin{figure}[h!]
%     \centering
%     \includegraphics[width=0.7\columnwidth]{figs/evals/coupleVPD.png}
%     \caption{Couple vp error}
%     \label{fig:coupleVPsim}
% \end{figure}

\subsubsection{Biological model}
% The biological~\cite{klipp2005systems} system \footnote{\url{https://ths.rwth-aachen.de/research/projects/hypro/biological-model-i/}} is modeled by the following ODE:
% \begin{align*}
%     \dot{x_1} &= -0.4\cdot x_1 + 5 \cdot x_3 \cdot x_4\\
%     \dot{x_2}
%     &=0.4\cdot x_1 -x_2\\
%     \dot{x_3}&=x_2-5\cdot x_3\cdot x_4\\
%     \dot{x_4} &= 5 \cdot x_5 \cdot x_6 - 5 \cdot x_3\cdot x_4\\
%     \dot{x_5}
%     &=5 \cdot - x_5 \cdot x_6 + 5 \cdot x_3\cdot x_4\\
%     \dot{x_6}&=0.5\cdot x_7 - 5\cdot x_5 \cdot x_6\\
%     \dot{x_7}&=-0.5\cdot x_7 + 5\cdot x_5\cdot x_6\\
% \end{align*}
We run the biological~\cite{klipp2005systems} system\footnote{\url{https://ths.rwth-aachen.de/research/projects/hypro/biological-model-i/}} with the following initial set:  $x_1(0),x_2(0),x_3(0),x_4(0),x_5(0),x_6(0),x_7(0), \in [0.99,1.01]$. In our experiments we use the following unsafe region: $x_4\leq 0.883 + 0.002 \cdot i$, with $i \in [1, 10]$. The linearized system contains 104 observables.
 Figure~\ref{fig:bioModelSim} compares behaviors of the original nonlinear and Koopman linearized systems.

%
% \begin{figure}[h!]
%     \centering
%     \includegraphics[width=0.7\columnwidth]{figs/evals/bioModel.png}
%     \caption{Bio data}
%     \label{fig:bioModelSim}
% \end{figure}
%
% \begin{figure}[h!]
%     \centering
%     \includegraphics[width=0.7\columnwidth]{figs/evals/bioModelD.png}
%     \caption{Bio error}
%     \label{fig:bioModelSim}
% \end{figure}

\subsubsection{Approximation Error} Although the goal of this work is not to investigate the Koopman linearization process, it is important to show that accurate approximations are plausible for the workflow to make sense.
Figure~\ref{fig:sims} showed a visual comparison of simulations and reachable sets for the Steam and Biological models, where the behaviors of the Koopman linearized systems appear close to the original nonlinear system.
A more detailed analysis of the linearization error for the dictionary of observables used is provided in Table~\ref{tbl:acc_error}.
Here, we run simulations from the corners and centers of the initial sets and compare the average and maximum errors at different points in time.
Although error generally grows as the simulation time increases, it remains acceptable within the time bound.
The maximum relative error observed is in the Roessler model at time $0.6 * T$, where the error reaches $5.76\%$.

\input{sections/accuracy_table}

Further investigation into error bounds for the Koopman linearization process is a potential topic of future research.
For general black-box systems where only data is provided, we do not expect guaranteed error bounds, although it may be possible to provide statistical guarantees.
When the observables are derived symbolically from the differential equations, guaranteed error bounds may be possible, although existing approaches for this are fairly pessimistic~\cite{forets2017explicit}.

%% file: sections/accuracy_table.tex
\begin{table}
  \caption{Simulation measurements of the Koopman linearization error up to the time bound $T$ show the relative error remains below a few percent.}
\label{tbl:acc_error}
% \bgroup
\setlength{\tabcolsep}{5pt}
\centering
\begin{tabular}{@{}lrlllll@{}}%
\toprule
\multicolumn{2}{r}{\textbf{Time} $\rightarrow$} & $\mathbf{T * 0.2}$ & $\mathbf{T * 0.4}$ & $\mathbf{T * 0.6}$ & $\mathbf{T * 0.8}$ & $\mathbf{T}$ \\
\midrule
\multirow{4}{*}{Coupled VP} & max abs error & 1.44E-03 & 3.73E-03 & 7.01E-03 & 1.47E-02 & 1.51E-02 \\
 & avg abs error & 6.06E-04 & 1.35E-03 & 2.90E-03 & 4.99E-03 & 5.06E-03 \\
 & max rel error & 1.46E-03 & 2.83E-03 & 4.36E-03 & 9.08E-03 & 9.39E-03 \\
 & avg rel error & 5.96E-04 & 1.01E-03 & 1.90E-03 & 3.23E-03 & 3.27E-03 \\ \hdashline[1pt/1pt]
\multirow{4}{*}{Biological} & max abs error & 3.28E-04 & 6.27E-04 & 4.33E-04 & 8.08E-04 & 1.18E-03 \\
 & avg abs error & 1.16E-04 & 2.11E-04 & 1.58E-04 & 2.75E-04 & 4.01E-04 \\
 & max rel error & 1.09E-04 & 1.99E-04 & 1.36E-04 & 2.53E-04 & 3.70E-04 \\
 & avg rel error & 3.82E-05 & 6.68E-05 & 4.97E-05 & 8.65E-05 & 1.26E-04 \\ \hdashline[1pt/1pt]
\multirow{4}{*}{Steam} & max abs error & 5.81E-04 & 5.73E-04 & 8.82E-04 & 1.13E-03 & 1.37E-03 \\
 & avg abs error & 2.89E-04 & 3.57E-04 & 3.41E-04 & 4.60E-04 & 6.47E-04 \\
 & max rel error & 4.80E-04 & 5.60E-04 & 1.03E-03 & 1.62E-03 & 2.43E-03 \\
 & avg rel error & 2.39E-04 & 3.56E-04 & 4.12E-04 & 6.81E-04 & 1.18E-03 \\ \hdashline[1pt/1pt]
\multirow{4}{*}{Roessler} & max abs error & 1.01E-01 & 9.54E-02 & 4.06E-01 & 3.91E-01 & 3.17E-01 \\
 & avg abs error & 5.20E-02 & 4.23E-02 & 1.76E-01 & 1.79E-01 & 2.36E-01 \\
 & max rel error & 1.04E-02 & 1.16E-02 & 5.76E-02 & 5.41E-02 & 3.70E-02 \\
 & avg rel error & 5.32E-03 & 5.08E-03 & 2.50E-02 & 2.47E-02 & 2.74E-02 \\ \bottomrule
\end{tabular}
\end{table}

%% file: sections/results.tex
\subsection{Results}

We run each of the models on a number of different instances.
As mentioned earlier, for each model we parameterize unsafe regions using a single parameter $i$ for each of the four systems.
In this section, we evaluate each of the proposed improvements to the direct encoding algorithm, as well as provide a comparison with other verification methods.
For space reasons, we do not present a detailed comparison of every optimization on every model instance, but instead elaborate on the overall trends.

%\begin{figure}
%\begin{subfigure}{.5\textwidth}
%    \centering
%    \includegraphics[width=1.\columnwidth]{figs/error/coupleVP.pdf}
%    \caption{Coupled Van der Pol oscillator. }
%    \label{fig:coupleVPerror}
%\end{subfigure}%
%\begin{subfigure}{.5\textwidth}
%    \centering
%    \includegraphics[width=1.1\columnwidth]{figs/error/steam.pdf}
%    \caption{Steam model.}
%    \label{fig:steamError}
%\end{subfigure}
%\caption{Evaluation results of the approximation error. We report minimal and maximum error computed for each time moment (grey lines are %intervals between these two values) and green points are average values.}
%\label{fig:error}
%\end{figure}

\subsubsection{Performance of Hyperplane Backpropagation} 
\begin{figure}[t]
    \centering
    \includegraphics[width=0.6\columnwidth]{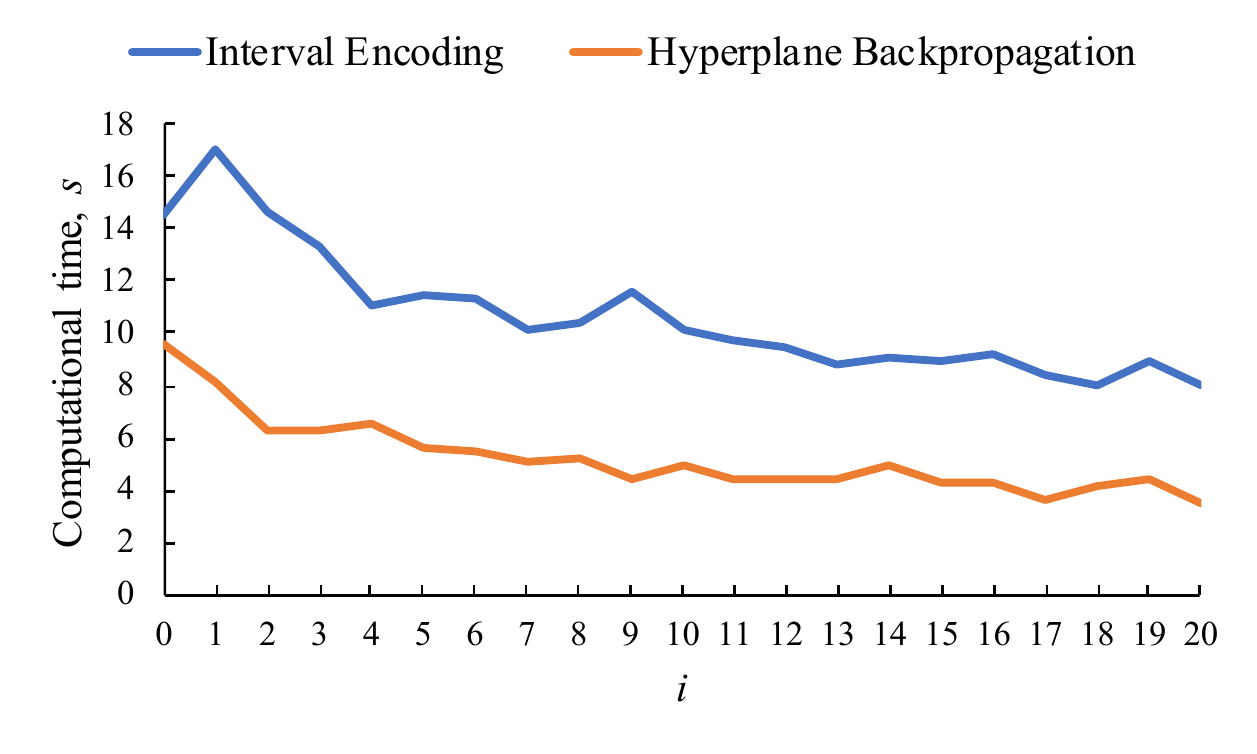}
    \caption{Evaluation results of Interval Encoding and  Hyperplane Backpropagation algorithms for the Roessler model for different values of $i$. Hyperplane Backpropagation is generally twice as fast for this problem.}
    \label{fig:roesslerBack}
\end{figure}

We compare the performance of the Hyperplane Backpropagation (Section~\ref{sec:hyback}) against the Interval Encoding algorithm (Section~\ref{sec:interval}).
Figure~\ref{fig:roesslerBack} shows a comparison with the Roessler model. We observe that the algorithm with backpropagation is around two times faster than the Interval Encoding algorithm for all problem instances.
In addition, the computational time gets smaller as $i$ is increased.
The main reason can be that we have a smaller time horizon when $i$ is large, because an unsafe state is reachable. We need to compute up to $t = 2.93$ for $i=1$ and up to $t = 2.81$ for $i=21$. We also call $\mathsf{dReal}$ less when $i$ is large for the same reason. For $i=1$, with the Interval Encoding Algorithm we call $\mathsf{dReal}$ 14 times, and only 8 times for Hyperplane Backpropagation.
For the last instance, $i = 21$, we call $\mathsf{dReal}$ 7 and 3 times for the two algorithms respectively, which leads to performance improvement.

%
% \begin{table}[]
% \centering
% \begin{tabular}{llllllllllllllllllllll}
% \textbf{} & \textbf{0} & \textbf{1} & \textbf{2} & \textbf{3} & \textbf{4} & \textbf{5} & 6 & 7 & 8 & 9 & 10 & 11 & 12 & 13 & 14 & 15 & 16 & 17 & 18 & 19 & 20 \\
% \textbf{No Refinements} & 14.549051 & 16.9901844 & 14.576705 & 13.2933422 & 11.0775624 & 11.4916068 & 11.2662201 & 10.0876502 & 10.3561571 & 11.5223285 & 10.1837157 & 9.77081002 & 9.43395016 & 8.75909158 & 9.02896962 & 8.92101154 & 9.24419052 & 8.44895562 & 7.94200451 & 8.96815872 & 7.98910123 \\
% \textbf{Back Refinement} & 9.53644536 & 8.09012507 & 6.34343102 & 6.23867106 & 6.52753977 & 5.59210678 & 5.43896457 & 5.10905263 & 5.19113156 & 4.43043937 & 5.00298478 & 4.44199668 & 4.37162024 & 4.41842399 & 4.89404565 & 4.2550717 & 4.2694661 & 3.6938949 & 4.20378334 & 4.37383057 & 3.50694755
% \end{tabular}
% \end{table}

\subsubsection{Performance of Zonotope Domain Splitting} We next evaluate Zonotope Domain Splitting (Algorithm~\ref{alg:domainContract}) on the Coupled Van der Pol oscillator.
We demonstrate the performance of the algorithm on two instances of the model: a safe case where $i = 4$ and an unsafe case where $i = 12$. We further evaluate using different values of $max\_level$.
We can observe that when the system is safe, Zonotope Domain Splitting with a large value of $max\_level$ generally benefits performance, whereas for $i = 12$ we see the opposite.
The explanation is that for the safe instance we can save calls to $\mathsf{dReal}$ with backpropagation and splitting together.
For $i = 12$, on the other hand, the system is unsafe and there are fewer steps where we can avoid calls to $\mathsf{dReal}$.
Performing splitting in this case is futile, as the overapproximation cannot prove safety of the system.

%\begin{figure}
%    \centering
%    \includegraphics[width=0.9\columnwidth]{figs/evals/coupleVPBin.pdf}
%    \caption{Evaluation results of the Zonotope Domain Splitting algorithm for the Coupled Van der Pol oscillator with $max\_level \in [0;5]$. Safe data series corresponds to $i=6.25$ and unsafe to $i = 8.25$.}
%    \label{fig:coupleBin}
%\end{figure}%

\begin{table}[b]
\setlength{\tabcolsep}{5pt}
\centering
\caption{Evaluation results of the Zonotope Domain Splitting algorithm for the Coupled Van der Pol oscillator with $max\_level \in [0, 5]$. Safe corresponds to the $i=4$ case and unsafe to the $i = 12$ case.}
\label{table:coupleBin}
\begin{tabular}{@{}lllllll@{}}%
\toprule
$max\_level$ & 0 & 1 & 2 & 3 & 4 & 5 \\
\midrule
\textbf{Safe} & 190.56 & 200.94 & 136.78 & 135.29 & 133.68 & 117.24 \\
\textbf{Unsafe} & 37.77 & 49.45 & 50.73 & 54.69 & 62.16 & 74.34 \\ \bottomrule
\end{tabular}
\end{table}

\subsubsection{Overall Algorithm Performance} 
\begin{figure}[t]
    \centering
    \includegraphics[width=0.8\columnwidth]{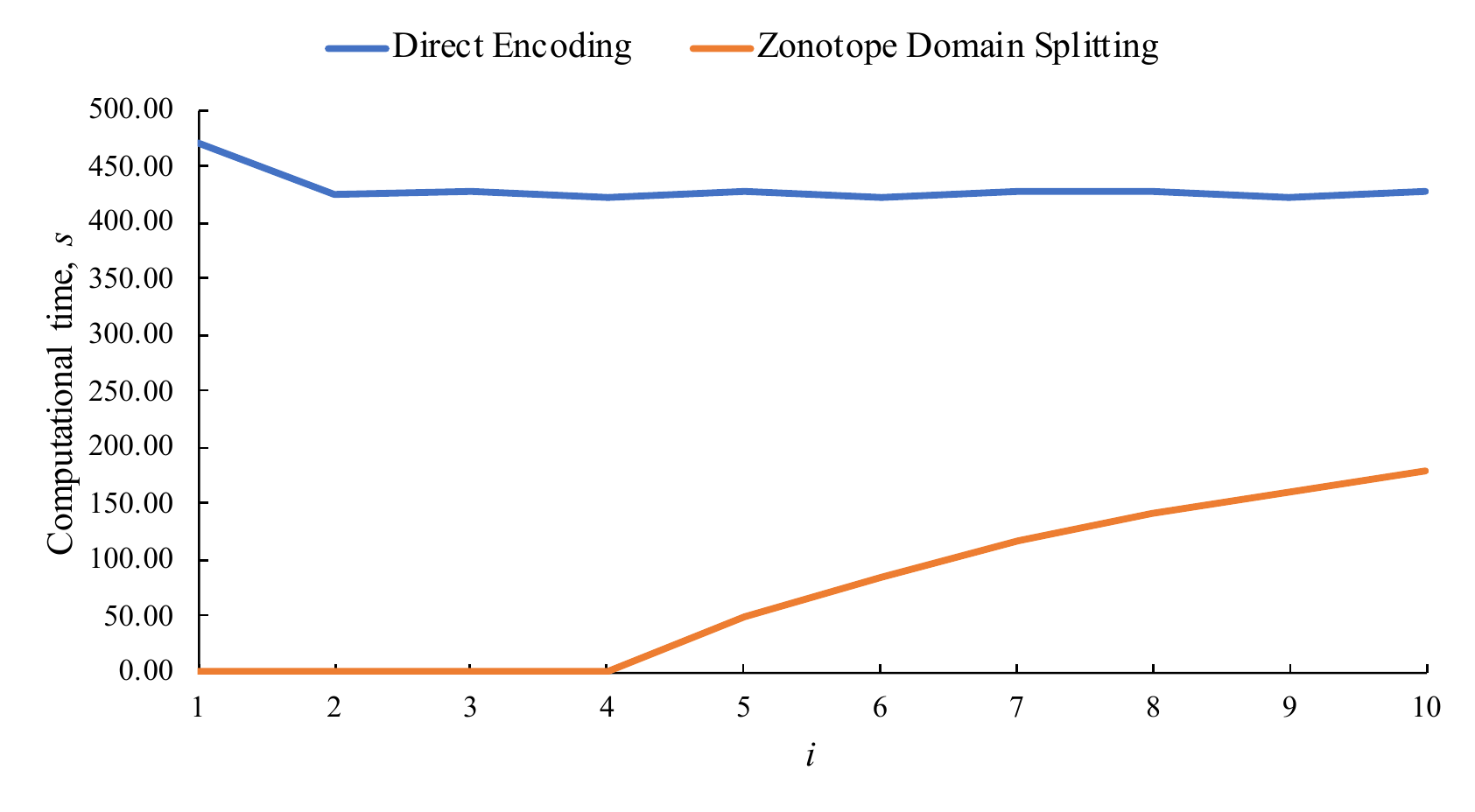}
    \caption{Comparison of Direct Encoding and Zonotope Domain Splitting on the Biological model. When $i$ is small, the optimized Zonotope Domain Splitting method verifies the system without calling $\mathsf{dReal}$, and is over 1000 times faster.}
    \label{fig:bioDreal}
\end{figure}

We next compare performance against the Direct Encoding algorithm (Section~\ref{sec:directdreal}), using the Biological model.
In Figure~\ref{fig:bioDreal}, for $i \in [1,4]$ we observe the computation time is close to zero and we are 1000 times faster than when we use Direct Encoding to verify the system.
We gain such a boost due to avoiding calls to $\mathsf{dReal}$ altogether, by only employing backpropagation.
Put another way, the overapproximation is sufficient to verify the system as safe when the reachable states are far from the unsafe set.
When $i \geq 5$ we can observe that the computational time for our approach increases due to $\mathsf{dReal}$ calls. However, since we still eliminate many unnecessary calls to $\mathsf{dReal}$, our algorithm still verifies the system $4-8$ times faster.
The runtime of the direct encoding method is much less sensitive to the $i$ parameter.

%\begin{table}[]
%\setlength{\tabcolsep}{2pt}
%\def\arraystretch{1.0}
%\centering
%\begin{tabular}{@{}rllllllllll@{}}%
%\toprule
%{\textit{$i \rightarrow$}} & 1& 2 & 3& 4& 5 & 6& 7& 8 & 9 & 10 \\
%\midrule
%\begin{tabular}{@{}l@{}}\textbf{Direct}\end{tabular} & 470.59 & 424.19 & 427.10 & 422.23 & 426.37 & 422.91 & 426.57 & 427.27 & 421.70 & 427.00 \\
%\begin{tabular}{@{}l@{}}\textbf{Zonotope}\end{tabular}& 0.59 & 0.43 & 0.43 & 0.42 & 49.92 & 84.76 & 117.59 & 141.97 & 159.83 & 179.25 \\ \bottomrule
%\end{tabular}
%\caption{Comparison of Direct Encoding and Zonotope Domain Splitting on the Biological model with $i \in [1,10]$. Computational times are presented in seconds.}
%\label{table:bioDreal}
%\end{table}

\subsubsection{Comparison with Other Methods}
We next summarize the verification time for each of the models using Zonotope Domain Splitting, Direct Encoding, the Flow*~\cite{chen2013flow} nonlinear reachability tool and the dReach tool~\cite{kong2015dreach}, which uses dReal to directly verify the original nonlinear systems.
The result are shown in Table~\ref{tbl:error}. The Flow* tool parameters were taken from the HyPro benchmark repository~\cite{schupp2017h}---developed by the same group as Flow*---ensuring the measured performance is not due to a poor choice of parameters.
%
%Thus, we assume that reachability configuration parameters are optimal.
%
The Zonotope Domain Splitting algorithm is up to 1000 times faster than both Flow* and Direct Encoding on many instances.
However, there are instances (\textit{e.g.} Steam model with $i=0$) where Flow* performs better.
Such cases outline directions for future research directions, for example trying to find ways to guide the search for the counterexample when splitting the zonotope domain, as opposed to always using the largest variable range.
The dReach tool timed out on all of the models, meaning that a verification result was not produced within two hours.
We confirmed the tool was being run correctly by severely reducing the analysis time bound until an output was produced.

% \textcolor{red}{This should be a table. Also run dReal if possible. Which value of i are you using? Also the the best performance for our algorithm. Also add unsafe intersection with FloW* (should slow it down), and parameters used for Flow* if possible}

\input{sections/comp_time_table}

%% file: sections/comp_time_table.tex
\begin{table}[t]
  \caption{Computational time (seconds) comparing Flow*, Direct Encoding and the Zonotope Domain Splitting. The dReach tool timed out on all models.}
\label{tbl:error}
% \bgroup
\setlength{\tabcolsep}{5pt}
\centering
\begin{tabular}{@{}lrlllcc@{}}%
\toprule
\multicolumn{2}{r}{\textit{$i$}} & Flow* & \begin{tabular}{@{}c@{}}Direct\end{tabular} & \begin{tabular}{@{}c@{}}Zono \end{tabular} \\
\midrule
\multirow{3}{*}{Coupled VP} & 1 & 251.11 & 788.45	 & 0.57 &  \\
 & 8 &  497.61 & 680.61	& 53.91 \\
 & 16 & 1665.16 & 557.24	 & 18.52 \\
 \hdashline[1pt/1pt]
\multirow{3}{*}{Biological} & 1 & 260.69 & 470.59	& 0.59  \\
 & 5  & 250.26 & 426.37 & 49.41\\
 & 10 & 238.56 & 427.00	& 179.25 \\\hdashline[1pt/1pt]
\multirow{3}{*}{Steam} & 0  & 61.06 & 197.08 & 182.62 \\
 & 5 & 285.20 & 59.53 & 37.27  \\
 & 10 & 77.68 & 29.21	& 18.52  \\ \hdashline[1pt/1pt]
\multirow{3}{*}{Roessler} & 0 & 55.28 & 181.06 & 9.53 \\
 & 10 & 78.33 & 177.92 & 5.01  \\
 & 20 & 55.29 & 174.63 & 3.50  \\
 \bottomrule
\end{tabular}
\end{table}

%% file: sections/related.tex
\section{Related Work}
\label{sec:related}

Domain contraction, the practice of narrowing the domain of possible solutions, has been used in solving constraint satisfaction over non-linear real arithmetic~\cite{benhamou2006continuous,franzle2006efficient,gao2010integrating,granvilliers2006algorithm}.
The same technique has been used in hybrid systems safety verification~\cite{ratschan2007safety} and checking for intersection of flow-pipes with guard sets for discrete transitions~\cite{althoff2012avoiding,chen2012taylor}.
While in these works, the domain contraction is performed in a branch-and-bound manner, we propagate constraints and use a zonotope intersection method for computing the reduced domain.
%
% The domain contraction for narrowing the sub-space of possible unsafe executions is similar in flavor to~\cite{chen2012taylor}.

Transforming a non-linear ODE into a linear ODE by performing change of variables has also been previously investigated~\cite{sankaranarayanan2012change}.
The main goal is to a) search for the change of variables transformation such that non-linear dynamical and hybrid systems become linear dynamical and hybrid systems and b) synthesize invariants for the linear system to prove safety.
The current work differs in two ways.
First, our method is purely data-driven and hence can also be applied to black-box systems.
Second, instead of synthesizing invariants, we prove safety by computing the flow-pipe of the high dimensional linear system.

% Another popular method for approximating nonlinear systems as linear systems with uncertainty is \emph{Hybridization}.
%
The process of approximating a nonlinear dynamical system as a piecewise linear system with uncertain inputs is called hybridization~\cite{asarin2003reachability,asarin2007hybridization,dang2010accurate,han2006reachability,bak2016scalable,li-et-al-2020:formats-2020}.
The inputs overapproximate the divergence between the linear and nonlinear dynamics in a subspace defined by the invariant of each mode in the hybrid system.
The first challenge in performing safety verification using hybridization is that it requires computing intersections of flow-pipes with the guards for discrete transitions, which might become expensive~\cite{girard2008zonotope}.
Some methods to avoid computing this intersections have been investigated~\cite{althoff2012avoiding,bak2016scalable}.
Secondly, for accurate hybridization, the number of modes in the hybrid system might be prohibitively large.
This paper investigates techniques for approximating a non-linear system as a high dimensional linear system for safety verification use observable variables, rather than linear approximations of the original nonlinear dynamics.
While high dimensional linearization using Koopman operator has been used to analyze controllability in~\cite{goswami2017global} and propagating the probability distribution of initial set in~\cite{matavalam2020data}, these works do not explicitly construct flow-pipes for proving safety of non-linear systems.

Finally, flow-pipe computation techniques for computing reachable set using Taylor Models~\cite{chen2013flow}, polynomial Zonotopes~\cite{althoff2015introduction}, sample trajectories~\cite{duggirala2015c2e2}, and SMT solvers~\cite{kong2015dreach} have all been investigated as potential techniques for nonlinear systems verification.
These methods work directly on the nonlinear differential equations, which create scalability challenges, whereas we use Koopman Operator linearization prior to analysis which can leverage scalable methods to compute reachable states for linear systems.

%% file: sections/conclusions.tex
\section{Conclusion}

Accurate reachability and verification of nonlinear dynamical systems is a grand challenge.
Many methods have been proposed for this problem, and this paper has provided a new avenue to verification based on Koopman operator linearization.
This process outputs a system of linear dynamics with nonlinear constraints on the initial state set.
As far as we are aware, this class of systems has not been considered before for formal set-based reachability analysis.
We have proposed and evaluated the first algorithm to solve such reachability problems, along with several optimizations that use overapproximation to reduce analysis time.
We have demonstrated that on some systems our method can outperform existing mature reachability tools, and we expect with further research both performance and accuracy can be improved.
Unique to our method, we can compute reachable states for black-box systems, as the Koopman linearization process uses data to create linear approximations in the space of observable functions.
In the future we plan to further optimize the method, as well as to create a user-friendly tool in order to participate in the annual nonlinear reachability competition~\cite{geretti2020arch}.

%% file: sections/acks.tex
\section{Acknowledgements}

This research was supported in part by the National Science Foundation (NSF) under grant number CNS 1935724 and by the Air Force Office of Scientific Research under award numbers \mbox{FA2386-17-1-4065} and \mbox{FA9550-19-1-0288}. Any opinions, findings, and conclusions or recommendations expressed in this material are those of the authors and do not necessarily reflect the views of the United States Air Force.